\documentclass{ws-mplb}

\usepackage{amssymb}
\usepackage{amsbsy}
\usepackage{amsmath}
\usepackage{graphicx}

\begin{document}

\def\Sign{\Sigma_{\rm n}} 
\def\Sigan{\Sigma_{\rm an}} 
\def\Signt{\tilde\Sigma_{\rm n}} 
\def\Sigant{\tilde\Sigma_{\rm an}} 
\def\Gn{G_{\rm n}}
\def\Gan{G_{\rm an}}
\def\beq{\begin{equation}}
\def\eeq{\end{equation}}
\def\bleq{\begin{eqnarray}}
\def\eleq{\end{eqnarray}} 
\newcommand{\Tr}{{\rm Tr}} 
\newcommand{\mean}[1]{\langle #1 \rangle}
\newcommand{\ie}{i.e. }
\newcommand{\eg}{e.g. }
\newcommand{\cc}{{\rm c.c.}} 
\def\half{\frac{1}{2}}
\def\p{{\bf p}} 
\def\q{{\bf q}}
\def\r{{\bf r}}
\def\w{\omega}
\def\inttau{\int_0^\beta d\tau}
\def\dtau{{\partial_\tau}} 
\def\eps{\epsilon}
\def\nablabf{\boldsymbol{\nabla}}
\def\llbrace{\left\lbrace}
\def\rrbrace{\right\rbrace} 
\def\dt{\partial_t} 
\def\lamb{\lambda}
\def\para{\parallel}

\markboth{N. Dupuis}{Infrared behavior of interacting bosons at zero temperature}

%
\catchline{}{}{}{}{}
%

\title{Infrared behavior of interacting bosons at zero temperature}

\author{\footnotesize N. Dupuis}

\address{Laboratoire de Physique Th\'eorique de la Mati\`ere Condens\'ee, 
CNRS - UMR 7600, \\ Universit\'e Pierre et Marie Curie, 4 Place Jussieu, 
75252 Paris Cedex 05, France \\
dupuis@lptmc.jussieu.fr}

\maketitle

\begin{history}
\received{(Day Month Year)}
\revised{(Day Month Year)}
\end{history}

\begin{abstract}
We review the infrared behavior of interacting bosons at zero temperature. After a brief discussion of the Bogoliubov approximation and the breakdown of perturbation theory due to infrared divergences, we show how the non-perturbative renormalization group enables to obtain the exact infrared behavior of the correlation functions. 
\end{abstract}


\section{Introduction}

Many of the predictions of the Bogoliubov theory of superfluidity\cite{Bogoliubov47} have been confirmed experimentally, in particular in ultracold atomic gases.\cite{Dalfovo99,Leggett01} Nevertheless a clear understanding of the infrared behavior of interacting bosons at zero temperature has remained a challenging theoretical issue for a long time. Early attempts to go beyond the Bogoliubov theory have revealed a singular perturbation theory plagued by infrared divergences due to the presence of the Bose-Einstein condensate and the Goldstone mode.\cite{Hugenholtz59,Gavoret64,Beliaev58} In the 1970s, Nepomnyashchii and Nepomnyashchii proved that the anomalous self-energy vanishes at zero frequency and momentum in dimension $d\leq 3$.\cite{Nepomnyashchii75} This exact result shows that the Bogoliubov approximation, where the linear spectrum and the superfluidity rely on a finite value of the anomalous self-energy, breaks down at low energy. As realized latter on,\cite{Nepomnyashchii78,Nepomnyashchii83} the singular perturbation theory is a direct consequence of the coupling between transverse and longitudinal fluctuations and reflects the divergence of the longitudinal susceptibility -- a general phenomenon in systems with a continuous broken symmetry.\cite{Patasinskij73} 

In this paper, we review the infrared behavior of interacting bosons. A more detailed discussion together with a comparison to the classical O($N$) model can be found in Ref.~\cite{Dupuis10}. In Sec.~\ref{sec_bosons_pt}, we briefly review the Bogoliubov theory and the appearance of infrared divergences in perturbation theory. We introduce the Ginzburg momentum scale $p_G$ signaling the breakdown of the Bogoliubov approximation. In Sec.~\ref{sec_bosons_nprg}, we discuss the non-perturbative renormalization group (NPRG) and show how it enables to obtain the exact infrared behavior of the normal and anomalous single-particle propagators without encountering infrared divergences.\cite{Castellani97,Pistolesi04,Wetterich08,Floerchinger08,Dupuis07,Dupuis09a,Dupuis09b,Sinner09,Sinner10} 

\section{Perturbation theory and breakdown of the Bogoliubov approximation}
\label{sec_bosons_pt}

We consider interacting bosons at zero temperature with the (Euclidean) action
\begin{equation}
S = \int dx \left[ \psi^*\left(\dtau-\mu - \frac{\nablabf^2}{2m}
  \right) \psi + \frac{g}{2} (\psi^*\psi)^2 \right] ,
\label{action}
\end{equation}
where $\psi(x)$ is a bosonic (complex) field, $x=(\r,\tau)$, and $\int dx=\inttau \int d^dr$. $\tau\in [0,\beta]$ is an imaginary time, $\beta\to\infty$ the inverse temperature, and $\mu$ denotes the
chemical potential. The interaction is assumed to be local in space and the
model is regularized by a momentum cutoff $\Lambda$. We consider a space dimension $d>1$.

Introducing the two-component field 
\begin{equation}
\Psi(p) = \left( \begin{array}{c} \psi(p) \\ \psi^*(-p)  \end{array} \right) , \quad 
\Psi^\dagger(p) = \bigl( \psi^*(p), \psi(-p) \bigr) 
\end{equation}
(with $p=(\p,i\w)$ and $\w$ a Matsubara frequency), the one-particle (connected) propagator becomes a $2\times 2$ matrix whose inverse in Fourier space is given by
\begin{equation}
\left( 
\begin{array}{cc} i\w + \mu -\eps_\p -\Sign(p) & - \Sigan(p) \\
 -\Sigan^*(p) & -i\w + \mu -\eps_\p -\Sign(-p)
\end{array}
\right) ,
\label{propa}
\end{equation}
where $\Sign$ and $\Sigan$ are the normal and anomalous self-energies, respectively, and $\eps_\p=\p^2/2m$. If we choose the order parameter $\mean{\psi(x)}=\sqrt{n_0}$ to be real (with $n_0$ the condensate density), then the anomalous self-energy $\Sigan(p)$ is real.

\subsection{Bogoliubov approximation} 
\label{subsec_bog}

The Bogoliubov approximation is a Gaussian fluctuation theory about the saddle point solution $\psi(x)=\sqrt{n_0}=\sqrt{\mu/g}$. It is equivalent to a zero-loop calculation of the self-energies,\cite{AGD_book,Fetter_book}
\begin{equation}
\Sign^{(0)}(p) = 2gn_0 , \quad
\Sigan^{(0)}(p) = gn_0 .
\end{equation}
This yields the (connected) propagators 
\begin{eqnarray}
G_{\rm n}^{(0)}(p) &=& -\mean{\psi(p)\psi^*(p)}_c = \dfrac{-i\w-\eps_\p-gn_0}{\w^2+E_\p^2} , \nonumber \\ 
G_{\rm an}^{(0)}(p) &=& -\mean{\psi(p)\psi(-p)}_c = \dfrac{gn_0}{\w^2+E_\p^2} ,
\end{eqnarray}
where $E_\p=[\eps_\p(\eps_\p+2gn_0)]^{1/2}$ is the Bogoliubov quasi-particle excitation energy. When $|\p|$ is larger than the healing momentum $p_c=(2gmn_0)^{1/2}$, the spectrum $E_\p\simeq \eps_\p+gn_0$ is particle-like, whereas it becomes sound-like for $|\p|\ll p_c$ with a velocity $c=\sqrt{gn_0/m}$. In the weak-coupling limit, $n_0\simeq \bar n$ ($\bar n$ is the mean boson density) and $p_c$ can equivalently be defined as $p_c=(2gm\bar n)^{1/2}$. In the Bogoliubov approximation, the occurrence of a linear spectrum at low energy (which implies superfluidity according to Landau's criterion) is due to $\Sigan(p=0)$ being nonzero. 

\subsection{Infrared divergences and the Ginzburg scale}
\label{subsec_bosons_ir}

The lowest-order (one-loop) correction $\Sigma^{(1)}$ to the Bogoliubov result $\Sigma^{(0)}$ is divergent for $d\leq 3$. Retaining only the divergent part, we obtain 
\beq
\Sign^{(1)}(p) \simeq \Sigan^{(1)}(p) \simeq - 2 \frac{g^4 n_0^3}{c^3} A_{d+1} \left(\p^2+\frac{\w^2}{c^2}\right)^{(d-3)/2} 
\label{sigma4a}
\eeq
if $d<3$ and 
\beq
\Sign^{(1)}(p) \simeq \Sigan^{(1)}(p) \simeq - \frac{g^4 n_0^3}{c^3} A_{4} \ln\left(\frac{p^2_c}{\p^2+\w^2/c^2}\right)  
\label{sigma4b}
\eeq
if $d=3$, where 
\beq
A_d = \llbrace 
\begin{array}{lcc}
- \frac{2^{1-d}\pi^{1-d/2}}{\sin(\pi d/2)} \frac{\Gamma(d/2)}{\Gamma(d-1)} & \mbox{if} & d<4 ,\\ 
\frac{1}{8\pi^2} & \mbox{if} & d=4 .
\end{array}
\right. 
\eeq
We can estimate the characteristic (Ginzburg) momentum scale $p_G$ below which the Bogoliubov approximation breaks down from the condition $|\Sign^{(1)}(p)| \sim \Sign^{(0)}(p)$ or  $|\Sigan^{(1)}(p)| \sim \Sigan^{(0)}(p)$ for $|\p|=p_G$ and $|\w|=cp_G$,
\begin{equation}
p_G \sim \left\lbrace 
\begin{array}{lcc}
(A_{d+1} gm p_c)^{1/(3-d)} & \mbox{if} & d<3 , \\
p_c \exp\left( - \frac{1}{A_4 gmp_c}\right) & \mbox{if} & d=3 .
\end{array}
\right.
\label{pG_est}
\end{equation}
This result can be rewritten as
\begin{equation}
p_G \sim \left\lbrace 
\begin{array}{lcc}
p_c (A_{d+1} \tilde g^{d/2})^{1/(3-d)} & \mbox{if} & d<3 , \\
p_c \exp\left( - \frac{1}{A_4 \sqrt{2} \tilde g^{3/2}}\right) & \mbox{if} & d=3 ,
\end{array}
\right.
\end{equation}
where 
\beq
\tilde g = gm \bar n^{1-2/d} \sim \left(\frac{p_c}{\bar n^{1/d}}\right)^2
\eeq
is the dimensionless coupling constant obtained by comparing the mean interaction energy per particle $g\bar n$ to the typical kinetic energy $1/m\bar r^2$ where $\bar r\sim \bar n^{-1/d}$ is the mean distance between particles.\cite{Petrov04} A superfluid is weakly correlated if $\tilde g\ll 1$, \ie $p_G\ll p_c\ll \bar n^{1/d}$ (the characteristic momentum scale $\bar n^{1/d}$ does however not play any role in the weak-coupling limit). In this case, the Bogoliubov theory applies to a large part of the spectrum where the dispersion is linear (\ie $|\p|\lesssim p_c$) and breaks down only at very small momenta  $|\p|\lesssim p_G\ll p_c$. When the dimensionless coupling $\tilde g$ becomes of order unity, the three characteristic momentum scales $p_G\sim p_c\sim \bar n^{1/d}$ become of the same order. The momentum range $[p_G,p_c]$ where the linear spectrum can be described by the Bogoliubov theory is then suppressed. We expect the strong-coupling regime $\tilde g\gg 1$ to be governed by a single characteristic momentum scale, namely $\bar n^{1/d}$.\cite{note2}

\subsection{Vanishing of the anomalous self-energy} 
\label{subsec_bosons_sigmaexact}

The exact values of $\Sign(p=0)$ and $\Sigan(p=0)$ can be obtained using the U(1) symmetry of the action, \ie the invariance under the field transformation $\psi(x)\to e^{i\theta}\psi(x)$ and $\psi^*(x)\to e^{-i\theta}\psi^*(x)$. On the one hand, the self-energies satisfy the Hugenholtz-Pines theorem,\cite{Hugenholtz59}
\beq
\Sign(p=0)-\Sigan(p=0)=\mu . 
\label{sigma6a}
\eeq
On the other hand, the anomalous self-energy vanishes,
\beq
\Sigan(p=0) = 0 .
\label{sigma6b}
\eeq 
The last result was first proven by Nepomnyashchii and Nepomnyashchii.\cite{Nepomnyashchii75,Sinner10,Dupuis10} It shows that the Bogoliubov theory, where the linear spectrum and the superfluidity rely on a finite value of the anomalous self-energy, breaks down at low energy in agreement with the conclusions drawn from perturbation theory (Sec.~\ref{subsec_bosons_ir}).

\section{The non-perturbative RG} 
\label{sec_bosons_nprg} 

The NPRG enables to circumvent the difficulties of perturbation theory and derive the correlation functions in the low-energy limit.\cite{Castellani97,Pistolesi04,Dupuis07,Dupuis09a,Dupuis09b,Wetterich08,Floerchinger08,Sinner09,Sinner10} The strategy of the NPRG is to build a family of theories indexed by a momentum scale $k$ such that fluctuations are smoothly taken into account as $k$ is lowered from the microscopic scale $\Lambda$ down to 0.\cite{Berges02,Delamotte07} This is achieved by adding to the action (\ref{action}) an infrared regulator term
\beq
\Delta S_k[\psi^*,\psi] = \sum_{p} \psi^*(p) R_k(p) \psi(p) .
\label{irreg}
\eeq
The main quantity of interest is the so-called average effective action
\beq
\Gamma_k[\phi^*,\phi] = - \ln Z_k[J^*,J] + \sum_{p} [J^*(p)\phi(p)+\cc] - \Delta S_k[\phi^*,\phi],
\eeq
defined as a modified Legendre transform of $-\ln Z_k[J^*,J]$ which includes the subtraction of $\Delta S_k[\phi^*,\phi]$. $J$ denotes a complex external source that couples linearly to the boson field $\psi$ and $\phi(x) = \mean{\psi(x)}$ is the superfluid order parameter. The cutoff function $R_k$ is chosen such that at the microscopic scale $\Lambda$ it suppresses all fluctuations, so that the mean-field approximation $\Gamma_\Lambda[\phi^*,\phi]=S[\phi^*,\phi]$ becomes exact. The effective action of the original model (\ref{action}) is given by $\Gamma_{k=0}$ provided that $R_{k=0}$ vanishes. For a generic value of $k$, the cutoff function $R_k(p)$ suppresses fluctuations with momentum $|\p|\lesssim k$ and frequency $|\w|\lesssim ck$ but leaves those with $|\p|,|\w|/c\gtrsim k$ unaffected ($c\equiv c_k$ is the velocity of the Goldstone mode). The dependence of the average effective action on $k$ is given by Wetterich's equation\cite{Wetterich93} 
\beq
\dt \Gamma_k[\phi^*,\phi] = \half \Tr\llbrace \dot R_k\left(\Gamma^{(2)}_k[\phi^*,\phi] + R_k\right)^{-1} \rrbrace ,
\label{rgeq}
\eeq
where $t=\ln(k/\Lambda)$ and $\dot R_k=\dt R_k$. $\Gamma^{(2)}_k[\phi^*,\phi]$ denotes the second-order functional derivative of $\Gamma_k[\phi]$. In Fourier space, the trace involves a sum over momenta and frequencies as well as the internal index of the $\phi$ field.

\subsection{Derivative expansion and infrared behavior} 
\label{subsec_bosons_de} 

Because of the regulator term $\Delta S_k$, the vertices $\Gamma^{(n)}_{k}(p_1,\cdots,p_n)$ are smooth functions of momenta and frequencies and can be expanded in powers of $\p_i^2/k^2$ and $\w_i^2/c^2k^2$. Thus if we are interested only in the low-energy properties, we can use a derivative expansion of the average effective action.\cite{Berges02,Delamotte07} In the following we consider the ansatz\cite{note1} 
\beq
\Gamma_k[\phi^*,\phi] = \int dx\Bigl[ \phi^*\Bigl(Z_{C,k}\dtau - V_{A,k}\partial^2_\tau - \frac{Z_{A,k}}{2m} \nablabf^2 \Bigr) \phi + \frac{\lamb_k}{2} (n-n_{0,k})^2 \Bigr] ,
\label{effaction} 
\eeq 
where $n=|\phi|^2$. $n_{0,k}$ denotes the condensate density in the equilibrium state. We have introduced a second-order time derivative term. Although not present in the initial average effective action $\Gamma_\Lambda$, we shall see that this term plays a crucial role when $d\leq 3$.\cite{Wetterich08,Dupuis07}

In a broken U(1) symmetry state with real order parameter $\phi=\sqrt{n_0}$, the normal and anomalous self-energies are given by 
\bleq
\Sigma_{k,{\rm n}}(p) &=& \mu + V_{A,k}\w^2 + (1- Z_{C,k})i\w 
- (1-Z_{A,k})\eps_\p + \lamb_k n_{0,k} , \\ 
\Sigma_{k,{\rm an}}(p) &=& \lamb_{k}n_{0,k} . \nonumber 
\eleq
These expressions imply the existence of a sound mode with velocity
\begin{equation}
c_k = \left( \frac{Z_{A,k}/2m}{V_{A,k}+Z_{C,k}^2/2\lambda_k n_{0,k}} \right)^{1/2} .
\label{cdef} 
\end{equation}
At the initial stage of the flow, $Z_{A,\Lambda}=Z_{C,\Lambda}=1$, $V_{A,\Lambda}=0$, $\lamb_\Lambda=g$   and $n_{0,\Lambda}=\mu/g$, which reproduces the results of the Bogoliubov approximation. A crucial property of the RG flow is that 
\beq
\lambda_k\sim k^{3-d}
\label{lambk1}
\eeq
vanishes with $k$ when $d\leq 3$. Eq.~(\ref{lambk1}) follows from the numerical solution of the RG equations, but can also be anticipated from the expected singular behavior of the longitudinal propagator.\cite{Dupuis10}

The parameters $Z_{A,k}$, $Z_{C,k}$ and $V_{A,k}$ can be related to thermodynamic quantities using Ward identities,\cite{Gavoret64,Pistolesi04,Dupuis09b}
\begin{eqnarray}
n_{s,k} &=& Z_{A,k} n_{0,k} = \bar n_k , \nonumber \\
V_{A,k} &=& - \frac{1}{2n_{0,k}} \frac{\partial^2 U_k}{\partial\mu^2}\biggl|_{n_{0,k}} ,\label{ward1} \\ 
Z_{C,k} &=& - \frac{\partial^2 U_k}{\partial n\partial\mu}\biggl|_{n_{0,k}} = \lambda_k \frac{dn_{0,k}}{d\mu} , \nonumber 
\end{eqnarray}
where $\bar n_k$ is the mean boson density and $n_{s,k}$ the superfluid density. Here we consider the effective potential $U_k$ as a function of the two independent variables $n$ and $\mu$. The first of equations~(\ref{ward1}) states that in a Galilean invariant superfluid at zero temperature, the superfluid density is given by the full density of the fluid.\cite{Gavoret64} Equations~(\ref{ward1}) also imply that the Goldstone mode velocity $c_k$ coincides with the macroscopic sound velocity,\cite{Gavoret64,Pistolesi04,Dupuis09b} \ie
\begin{equation}
\frac{d\bar n_k}{d\mu} = \frac{\bar n_k}{mc_k^2} .
\end{equation}
Since thermodynamic quantities, including the condensate ``compressibility''  $dn_{0,k}/d\mu$ should remain finite in the limit $k\to 0$, we deduce from (\ref{ward1}) that $Z_{C,k} \sim \lambda_k \sim k^{3-d}$ vanishes in the infrared limit, and 
\begin{equation}
\lim_{k\to 0} c_k = \lim_{k\to 0} \left( \frac{Z_{A,k}}{2mV_{A,k}} \right)^{1/2} .
\label{velir}
\end{equation}
Both $Z_{A,k}=\bar n_k/n_{0,k}$ and the macroscopic sound velocity $c_k$ being finite at $k=0$, $V_{A,k}$ (which vanishes in the Bogoliubov approximation) takes a non-zero value when $k\to 0$. 

\begin{figure}
\centerline{\includegraphics[width=6cm,clip]{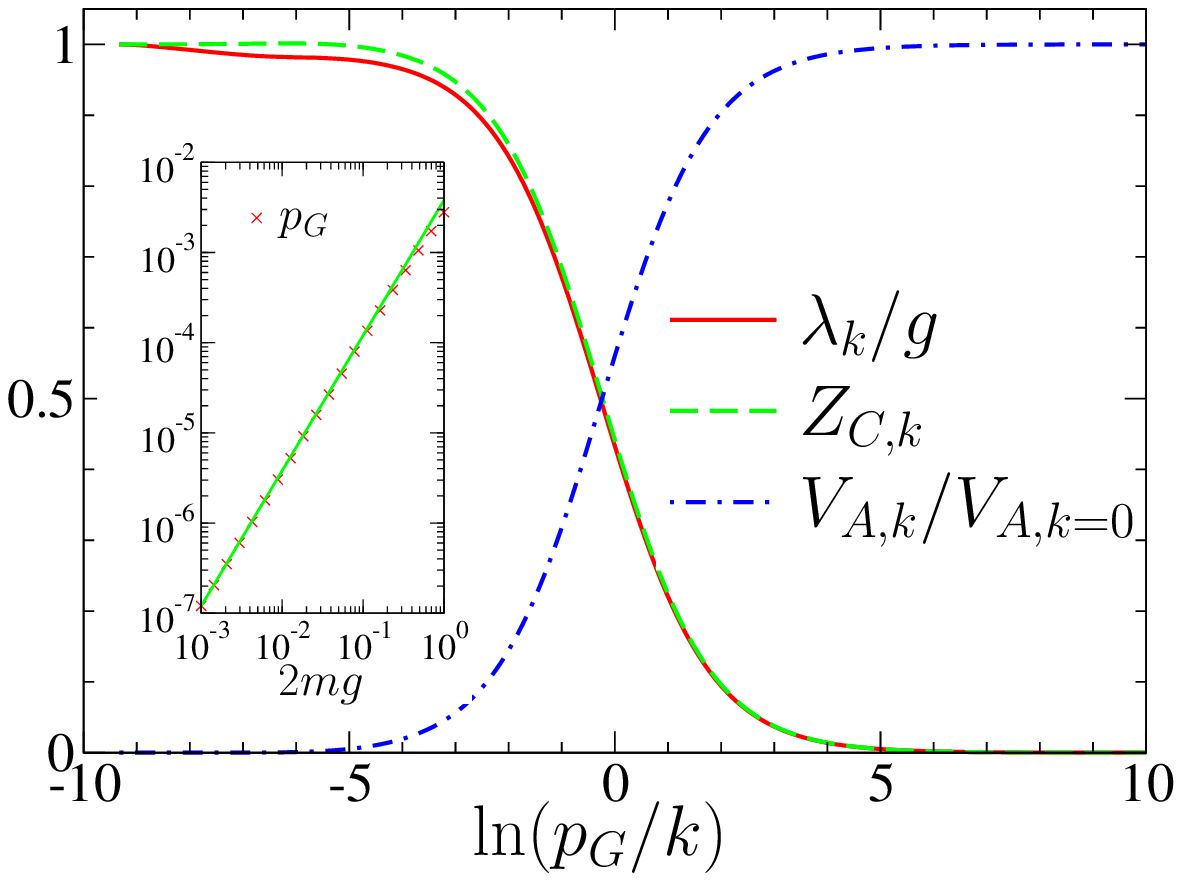}
\includegraphics[width=6cm,clip]{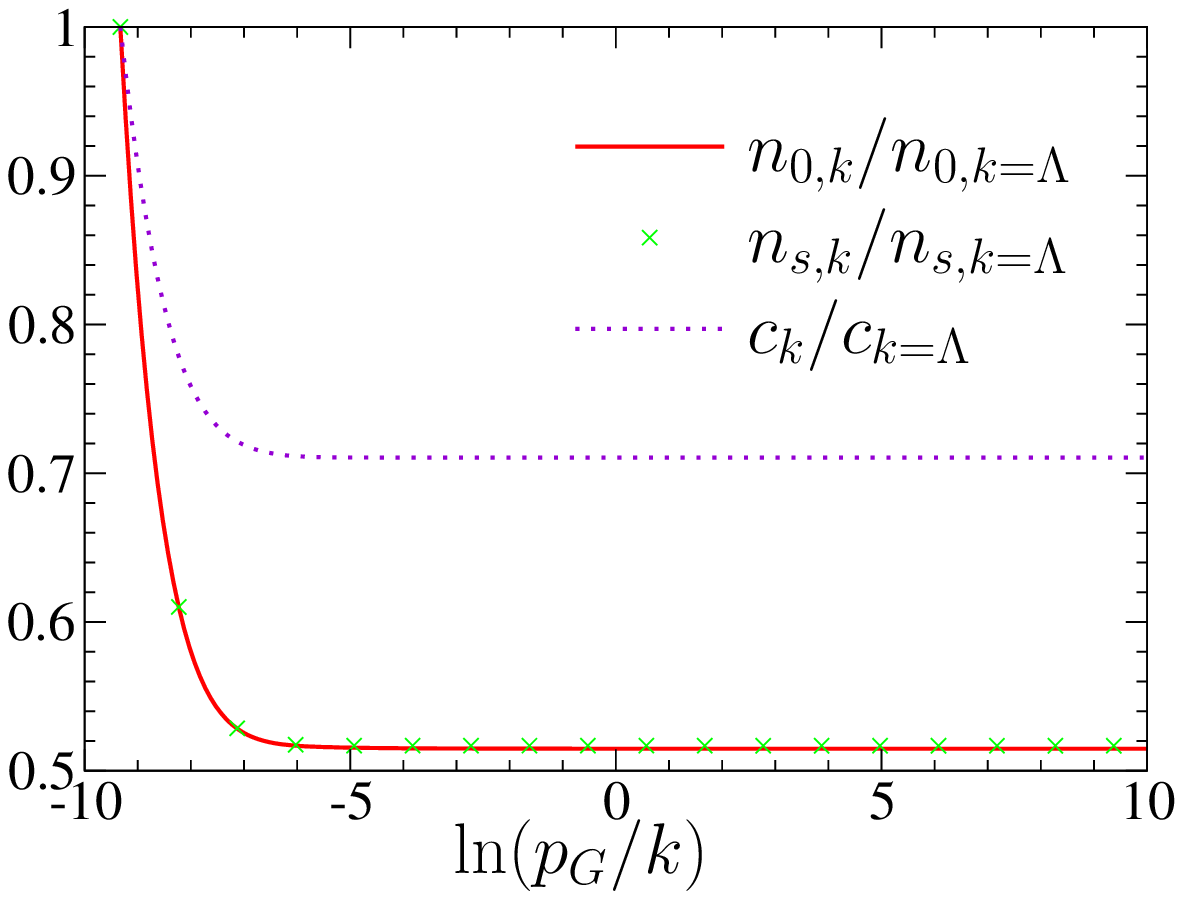}}
\caption{ (Color online) (Left panel) $\lamb_k$, $Z_{C,k}$ and $V_{A,k}$ vs $\ln(p_G/k)$ where $p_G=\sqrt{(gm)^3\bar n}/4\pi$ for $\bar n=0.01$, $2mg=0.1$ and $d=2$ [$\ln(p_G/p_c)\simeq -5.87$]. The inset shows $p_G$ vs $2mg$ obtained from the criterion $V_{A,p_G}=V_{A,k=0}/2$ [the Green solid line is a fit to $p_G\sim (2mg)^{3/2}$]. (Right panel) Condensate density $n_{0,k}$, superfluid density $n_{s,k}$ and Goldstone mode velocity $c_k$ vs $\ln(p_G/k)$.} 
\label{fig_boson_flow} 
\end{figure}

The suppression of $Z_{C,k}$, together with a finite value of $V_{A,k=0}$ shows that the average effective action (\ref{effaction}) exhibits a ``relativistic'' invariance in the infrared limit and therefore becomes equivalent to that of the classical O(2) model in dimensions $d+1$. In the ordered phase, the coupling constant of this model vanishes as $\lamb_k\sim k^{4-(d+1)}$,\cite{Dupuis10} which agrees with (\ref{lambk1}). For $k\to 0$, the existence of a linear spectrum is due to the relativistic form of the average effective action (rather than a non-zero value of $\lambda_k n_{0,k}$ as in the Bogoliubov approximation).

To obtain the $k=0$ limit of the propagators (at fixed $p$), one should in principle stop the flow when $k\sim \sqrt{\p^2+\w^2/c^2}$.\cite{Dupuis09b} Since thermodynamic quantities are not expected to flow in the infrared limit, they can be approximated by their $k=0$ values. Making use of the Ward identities~(\ref{ward1}), we deduce the exact infrared behavior of the normal and anomalous propagators (at $k=0$),\cite{Dupuis09b}
\bleq
G_{\rm n}(p) &=& - \frac{n_0mc^2}{\bar n} \frac{1}{\w^2+c^2\p^2}
- \frac{mc^2}{\bar n} \frac{dn_0}{d\mu} \frac{i\w}{\w^2+c^2\p^2} - \half G_{\para}(p) , \nonumber \\
G_{\rm an}(p) &=& \frac{n_0mc^2}{\bar n} \frac{1}{\w^2+c^2\p^2} - \half G_{\para}(p) ,  
\label{de3}
\eleq
where
\begin{equation}
G_{\para}(p) = \frac{1}{2n_0C (\w^2+c^2\p^2)^{(3-d)/2}} 
\label{de4}
\end{equation}
is the propagator of the longitudinal fluctuations. The constant $C$ follows from the replacement $\lambda_k\to C(\w^2+c^2\p^2)^{(3-d)/2}$. The leading terms in (\ref{de3}) agree with the results of Gavoret and Nozi\`eres.\cite{Gavoret64} The contribution of the diverging longitudinal correlation function was first identified by Nepomnyashchii and Nepomnyashchii.\cite{Nepomnyashchii78,Nepomnyashchii83}

\subsection{RG flows}
\label{subsec_bosons_num} 

The conclusions of the preceding section can be obtained more rigorously from the RG equation (\ref{rgeq}) satisfied by the average effective action. The flow of $\lamb_k$, $Z_{C,k}$ and $V_{A,k}$ is shown in Fig.~\ref{fig_boson_flow} for a two-dimensional system in the weak-coupling limit. We clearly see that the Bogoliubov approximation breaks down at a characteristic momentum scale $p_G\sim \sqrt{(gm)^3\bar n}$. In the Goldstone regime $k\ll p_G$, we find that both $\lamb_k$ and $Z_{C,k}$ vanish linearly with $k$ in agreement with the conclusions of Sec.~\ref{subsec_bosons_de}. Furthermore, $V_{A,k}$ takes a finite value in the limit $k\to 0$ in agreement with the limiting value (\ref{velir}) of the Goldstone mode velocity. Figure~\ref{fig_boson_flow} also shows the behavior of the condensate density $n_{0,k}$, the superfluid density $n_{s,k}=Z_{A,k}n_{0,k}$ and the velocity $c_k$. Since $Z_{A,k=0}\simeq 1.004$, the mean boson density $\bar n_k=n_{s,k}$ is nearly equal to the condensate density $n_{0,k}$. Apart from a slight variation at the beginning of the flow, $n_{0,k}$, $n_{s,k}=Z_{A,k}n_{0,k}$ and $c_k$ do not change with $k$. In particular, they are not sensitive to the Ginzburg scale $p_G$. This result is quite remarkable for the Goldstone mode velocity $c_k$, whose expression (\ref{cdef}) involves the parameters $\lamb_k$, $Z_{C,k}$ and $V_{A,k}$, which all strongly vary when $k\sim p_G$. These findings are a nice illustration of the fact that the divergence of the longitudinal susceptibility does not affect local gauge invariant quantities.\cite{Pistolesi04,Dupuis09b}

\section{Conclusion} 

Interacting bosons at zero temperature are characterized by two momentum scales: the healing (or hydrodynamic) scale $p_c$ and the Ginzburg scale $p_G$. $p_G$ sets the scale at which the Bogoliubov approximation breaks down. For momenta $|\p|\ll p_c$, it is possible to use a hydrodynamic description in terms of density and phase variables. This description allows one to compute the correlation functions without encountering infrared divergences.\cite{Popov79,Dupuis10} In this paper, we have reviewed another approach, based on the NPRG, which enables to describe the system at all energy scales and yields the exact infrared behavior of the single-particle propagator. A nice feature of the NPRG is that it can be used to study models of strongly-correlated bosons such as the Bose-Hubbard model.\cite{Rancon10} 


\begin{thebibliography}{10}
\bibitem{note1} One could consider a more general ansatz where $Z_{A,k}(n)$, $Z_{C,k}(n)$ and $V_{A,k}(n)$ are functions of the condensate density $n$, and the full effective potential $U_k(n)$ is retained rather than expanded to quadratic order about $n_{0,k}$. In the superfluid phase, $U_k(n)$ is convex but otherwise the conclusions regarding the infrared behavior are unchanged.
\end{thebibliography}

\end{document}